# A RADIOMICS APPROACH TO ANALYZE CARDIAC ALTERATIONS IN HYPERTENSION

*Irem Cetin[†], Steffen E. Petersen[‡], Sandy Napel[ǵ]*
*Oscar Camara[†] Miguel Ángel González Ballester[†§], Karim Lekadir[†]*

[†]BCN MedTech, Universitat Pompeu Fabra, Barcelona, Spain
[§] Catalan Institution for Research and Advanced Studies (ICREA), Barcelona, Spain
[‡]William Harvey Research Institute, Queen Mary University of London, London, UK
[ǵ] Department of Radiology, School of Medicine, Stanford University, Stanford, USA

## ABSTRACT

Hypertension is a medical condition that is well-established as a risk factor for many major diseases. For example, it can cause alterations in the cardiac structure and function over time that can lead to heart related morbidity and mortality. However, at the subclinical stage, these changes are subtle and cannot be easily captured using conventional cardiovascular indices calculated from clinical cardiac imaging. In this paper, we describe a radiomics approach for identifying intermediate imaging phenotypes associated with hypertension. The method combines feature selection and machine learning techniques to identify the most subtle as well as complex structural and tissue changes in hypertensive subgroups as compared to healthy individuals. Validation based on a sample of asymptomatic hearts that include both hypertensive and non-hypertensive cases demonstrate that the proposed radiomics model is capable of detecting intensity and textural changes well beyond the capabilities of conventional imaging phenotypes, indicating its potential for improved understanding of the longitudinal effects of hypertension on cardiovascular health and disease.

*Index Terms*— Hypertension, cardiovascular disease, radiomics, UK Biobank, machine learning.

## 1. INTRODUCTION

Hypertension is a major risk factor for cardiovascular diseases and events [1]. Approximately 77% of people who have a first stroke and 70% of people who have a first heart attack have hypertension [2]. This condition, while not directly linked to the heart, can induce longtidudinal alterations in the heart over a long period well before symptoms of cardiovascular disease develop. Eventually, this can lead to major cardiac diseases such as heart failure and left ventricular hypertrophy [3]. It is therefore of paramount importance to identify individuals at risk of developing hypertension-related diseases at early stage to apply preventive and corrective measures. However, existing cardiovascular imaging indices are not suitable to identify the subtle changes induced by hypertension at the subclinical stage. Furthermore, there is a major knowledge gap as to which perturbations take place in the heart over time in hypertensive patients and which lead to full-blown cardiovascular remodeling and dysfunctions.

In this paper, we propose a radiomics approach for deeper imaging phenotyping of cardiovascular alterations occurring in hypertensive patients with healthy hearts. To the best of our knowledge, this is the first radiomics based method and study dedicated to the quantification of subtle cardiac changes at the subclinical stage in asymptomatic individuals. The radiomics concept has been developed and exploited mostly in oncology with promising results for tumor classification and treatment planning [4, 5], while their usage in other clinical domains such as cardiology is only recent [6, 7]. In this paper, we analyze a large pool of advanced imaging features describing a range of shape, size, intensity and textural properties of the cardiac structures and tissues. By combining feature selection and supervised classification techniques, the goal is to identify a set of radiomic features that best discriminate between the normal and hypertensive cases in healthy hearts, well beyond the capabilities of conventional clinical indices such as ejection fraction. The proposed radiomics model is trained and validated on newly acquired datasets from the UK Biobank.

## 2. MATERIALS AND METHODS

### 2.1. Data description and segmentations

The UK Biobank is a large-scale population health resource (500,000 individuals) aimed at enhancing biomedical research like never before, ultimately improving prevention, diagnosis and treatment of a wide range of serious and life threatening illnesses such as heart disease. The UK Biobank holds an exceptional amount of data which includes biomedical data, physical measures, accelerometry, multimodal imaging including abdominal, brain and cardiac magnetic resonance imaging (MRI) scans as well as whole body DXA imaging, genome-wide genotyping and longitudinal follow-up for a wide range of health-related outcomes [8]. For this paper, 200 cardiac cine-MRI images were randomly selected, including 100 hypertensive patients and 100 cases without hypertension. Both subgroups have no evidence of cardiovascular disease. The cardiac images were acquired with 1.5 Tesla scan (MAGNETOM Area, Syngo Platform VD13A, Siemens Healthcare, Erlangen, Germany) and have an in-plane resolution of 1.8x1.8 $mm^2$, a slice thickness of 8.0 $mm$ and a slice gap of 2 $mm$. Manual annotation of the images was undertaken by our clinical collaborator (S.E. Pe- tersen), resulting in a segmentation of the left ventricle (LV), myocardium (MYO) and right ventricle (RV) boundaries [9].

**2.2. Radiomics feature calculation**

The goal of this work is to analyze a sufficiently large and diverse radiomics features such that relevant cardiac alterations due to hypertension can be captured. Concretely, we calculate a total of 686 radiomic features describing a range of shape, size, intensity or textural characteristics of the cardiac sub-structures and tissues (LV, MYO, RV). These include:

- Shape features will capture geometrical alterations in the cardiac structures, while size features will measure global and localized remodelling or dilatation/hypertrophy due to disease. The main shape/size radiomics include sphericity, compactness, elongation, ratios, diameters and main axes. In this paper, these estimated shape/size radiomics will be used to identify morphological remodeling occurring in hypertensive but not in non-hypertensive individuals.
- First-order intensity statistics will inform on the distribution of the gray level values in the cardiac tissues, without focusing on their spatial relationships. These include simple measures such as the mean intensity in a particular region of the tissue or the standard deviation, as well as more advanced measures such as skewness, uniformity or entropy.
- Advanced textural features will measure changes in the spatial relationships, local contrasts and tissue homogeneity within the different cardiac structures. These radiomic features can be beneficial, for example, to capture potential alterations in the trabeculae, papillary muscles and fibrosis in hypertensive vs. non-hypertensive subgroups. Different texture methods are included: Gray Level Co-occurrence Matrix (GLCM), Gray Level Run Length Matrix (GLRLM), Neighboring Gray Tone Difference Matrix (NGTDM) and Fractal Dimension (FD).

**2.3. Radiomic feature selection**

Our hypothesis in this work is that the most subtle changes in cardiac morphology, tissue or local texture due to hypertension, even in initial asymptomatic stages, will be captured by some of the multi-type radiomics, thus leading to new imaging phenotypes that could be used for early diagnosis of hypertension related subclinical heart disease. We expect that not all of the radiomics markers will be relevant for this analysis. Instead, a radiomic feature selection is necessary to select those features that specifically deviate from normality in the presence of hypertension.

In this work, we do this by combining sequential forward feature selection (SFFS) [10] with radiomics-based classification to identify the most relevant radiomics features as those that best classify hypertensive vs. non-hypertensive hearts in an SVM model.

Within a cross-validation scheme, the SFFS technique will enable to select sequentially, one at a time, the radiomic features that improve the overall classification of hypertensive vs. non-hypertensive individuals. The very first radiomic feature to be selected with this procedure is the one that best separate hypertensive and normal cases among all radiomic features. Sequentially, new radiomic features are added to provide complementary evidence for the classification and description of the hypertensive imaging phenotypes.

In this method, SVM is chosen as the underlying classification model due its well-known performance when classifying image data, in particular in the case of small sample size. An SVM model corresponds to a separation of the examples by the hyperplanes that have the largest distance to the nearest training-data point of any class (so-called functional margin). This ensures that the examples belonging to the different classes, in our cases hypertensive vs. normal hearts are separated as clearly as possible.

At the end of the procedure, radiomic features that have similar or overlapping distributions between the classes of interest are ignored, while those that contribute to the SVM classification of hypertensive and normal hearts are included within the final set of optimal radiomic features, indicating their relevance for describing hypertension related changes in asymptomatic hearts.

**3. RESULTS**

In this study, 10-fold cross validation tests are performed to select the optimal radiomic feature that best separate hypertensive and non-hypertensive hearts. Note that after application of the proposed feature selection method, the classification accuracy, measured as the number of correct classification divided by the number of cases, reaches 0.8. This confirms the hypothesis that hypertension does indeed alters the values of radiomics features even at the subclinical stage. This is further illustrated in Figure 1, which shows that using conventional imaging phenotypes of cardiovascular function, (i.e. ejection fractions, stroke volumes and volumes of left and right ventricles at end-diastolic and end-systolic time frames) results in a low classification of the normal and hy-pertension subgroups, with an area under curve (AUC) score of 0.62 ±0.09. This is an expected result as the hypertensive individuals are asymptomatic with normal cardiovascular structure and functions as evaluated by the clinicians. Instead, the proposed radiomics model significantly improves classification with AUC score of 0.76 0.13.

After demonstrating the relevance of the radiomics approach for discriminating the hypertension and normal subgroups, Table 1 lists the selected radiomics features, which sum to 11 features. It can be seen that all selected features are intensity and texture radiomics, which indicate that the main changes due to hypertension are in the actual tissues rather than the geometry and size of the ventricles or myocardium. This also explains the inability of conventional indices to characterize changes due to hypertension as these typically focus on the quantification of cardiac structure and function only. It is important to note that the selected radiomic features cover both end-diastole and end-systole (second column), as well as all three cardiac substructures (LV, MYO, RV - third column). Furthermore, The fourth column of the table lists the classification scores when using each feature alone, showing they do not separate well between the two subgroups, with the accuracy varying between 0.415 (Inverse difference moment nor-malized) and 0.635 (Gray level non-uniformity). This confirms that the hypertension related changes are indeed small and subtle. It is the combination of all the features into a radiomic signature, describing multiple co-occurring changes in the heart, that is best suited for optimal classification reaching a score of 0.8.

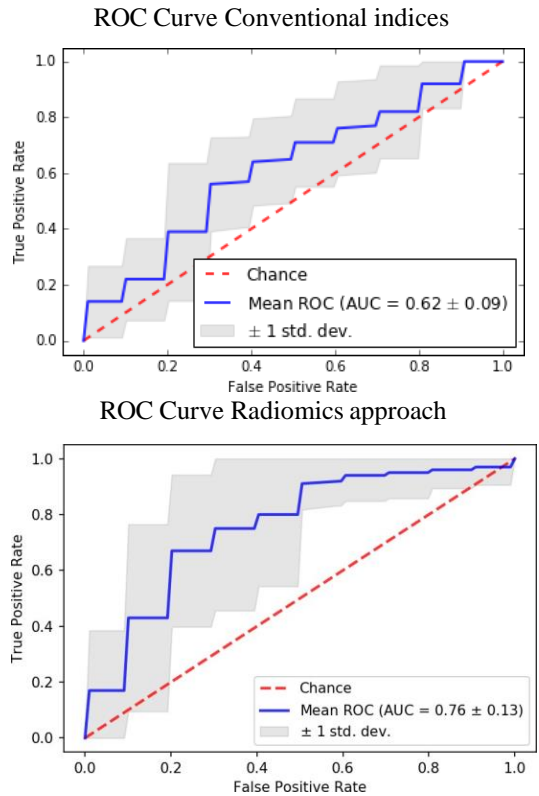

**Fig. 1**. ROC curves using the proposed method with selected radiomics features (top) and conventional imaging phenotypes (bottom).

**Table 1**. List of 11 radiomics features selected by the proposed method for discriminating the hearts of hypertensive and normal individuals. Alone: Classification accuracy using only this feature. W/O: Accuracy when removing the feature. ED: end-diastolic. ES: end-systolic.

| Name | Frame | Structure | Alone | W/O |
|---|---|---|---|---|
| Homogeneity 1 | ES | LV | 0.495 | 0.77 |
| Inverse variance | ES | LV | 0.55 | 0.685 |
| Inverse difference moment normalized | ED | MYO | 0.415 | 0.74 |
| Sum of squares | ED | MYO | 0.485 | 0.765 |
| Large area emphasis | ED | LV | 0.455 | 0.785 |
| Zone entropy | ED | LV | 0.56 | 0.725 |
| Large area low gray level emphasis | ED | RV | 0.485 | 0.765 |
| Short run emphasis | ES | RV | 0.505 | 0.79 |
| Long run emphasis | ED | MYO | 0.555 | 0.795 |
| Coarseness | ED | MYO | 0.62 | 0.76 |
| Gray level non-uniformity | ES | MYO | 0.635 | 0.71 |

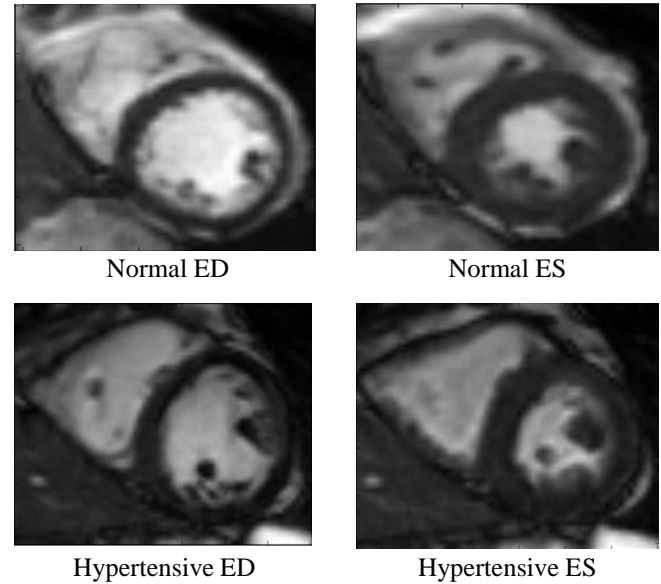

Normal ED     Normal ES

Hypertensive ED     Hypertensive ES

**Fig. 2**. Images of hypertensive and normal cases with the same LVEF values and different radiomics signature.

Finally, to further illustrate the benefit of the radiomics approach, Figure 2 shows a comparison between two hearts corresponding to normal (above) and hypertensive (bottom) cases, respectively. Both hearts look visually normal and furthermore they have the same left ventricular ejection fraction (LVEF) value (60%), indicating normal cardiac functions in both cases. In contrast, as shown in Table 2, the proposed radiomic signature enables to show clear differences between the two cases in the radiomic space. Several of the normalized radiomic values (using z-normalization) indicate differences between the values in the normal and hypertensive cases, such as for the large area emphasis (-1.31 vs. 0.15), short run emphasis (-2.21 vs. 0.32) and gray level non-uniformity (-2.19 vs. 1.42). The obtained results show the existence of textural differences between hypertensive and normal subgroups that cannot be captured by using clinical conventional indices such as EF or using visual examination.

Table 2. Original and normalized radiomics values for the two cases of Figure 2.

|  | Original | | Normalized | |
| --- | --- | --- | --- | --- |
|  | Normal | Hypertensive | Normal | Hypertensive |
| Homogeneity 1 | 0.543 | 0.546 | -0.199 | -0.133 |
| Inverse variance | 0.395 | 0.390 | 0.838 | 0.644 |
| **Inverse difference moment normalized** | 0.981 | 0.992 | **-1.3880** | **1.386** |
| Sum of squares | 0.706 | 0.936 | -0.467 | 0.161 |
| **Large area emphasis** | **196** | **6910** | **-1.315** | **0.151** |
| Zone entropy | 5.79 | 5.62 | 0.910 | 0.089 |
| Large area low gray level emphasis | 5 | 100 | -0.885 | -0.532 |
| **Short run emphasis** | 0.913 | 0.831 | **2.217** | **-0.325** |
| Long run emphasis | 1.76 | 2.80 | -1.692 | -0.401 |
| **Coarseness** | 0.00968 | 0.00203 | **7.761** | **-1.090** |
| **Gray level non-uniformity** | **298** | **1670** | **-2.192** | **1.427** |

## 4. DISCUSSION AND CONCLUSIONS

To the best of our knowledge, this is the first radiomics study performed to identify subclinical changes and quantify intermediate phenotypes associated with hypertension in otherwise healthy hearts. The results indicate that the main changes are in the cardiac textures and tissues, which explains the inability of conventional imaging indices, which focus on structural and functional quantification, to identify these alterations. This paper shows the promise of the proposed radiomics approach for analyzing subtle and more complex effects of non-cardiac risk factors of heart disease such as hypertension. Future work includes clinical interpretation of the results (e.g. fibrosis formation), as well as application to other risk factors such as diabetes and cholesterol effects.

## 5. ACKNOWLEDGEMENTS

This work is partly funded by the H2020 euCanSHare project (grant agreement 825903), and partly by the QUAES Foundation Chair for Computational Technologies for Healthcare. KL is funded by a Ramón y Cajal grant, MINECO, Spain.

## 6. REFERENCES

[1] S. E. Kjeldsen, "Hypertension and cardiovascular risk: General aspects," *Pharmacol. Res.*, vol. 129, 2018.

[2] R. Merai, C. Siegel, M. Rakotz, P. Basch, J. Wright, B. Wong, and P. Thorpe, "CDC Grand Rounds: A Public Health Approach to Detect and Control Hypertension," *MMWR Morb. Mortal. Wkly. Rep.*, vol. 65, 2016.

[3] W. S. Aronow, "Hypertension and left ventricular hypertrophy," *Ann Transl Med*, vol. 5, no. 15, 2017.

[4] C. Parmar, P. Grossmann, J. Bussink, P. Lambin, and H. J. Aerts, "Machine Learning methods for Quantitative Radiomic Biomarkers," *Sci Rep*, vol. 5, 2015.

[5] A. Kotrotsou, P. O. Zinn, and R. R. Colen, "Radiomics in Brain Tumors: An Emerging Technique for Characterization of Tumor Environment," *Magn Reson Imaging Clin N Am*, vol. 24, 2016.

[6] I. Cetin, G. Sanroma, S. E. Petersen, S. Napel, O. Camara, M-A G. Ballester, and K. Lekadir, "A radiomics approach to computer-aided diagnosis with cardiac cine-mri," in *Statistical Atlases and Computational Models of the Heart*, 2018.

[7] D. Dey and F. Commandeur, "Radiomics to identify high-risk atherosclerotic plaque from computed tomography," vol. 10.

[8] C. Sudlow et al., "UK biobank: an open access resource for identifying the causes of a wide range of complex diseases of middle and old age," *PLoS Med.*, vol. 12, 2015.

[9] S E. Petersen, P M. Matthews, J M. Francis, M D. Robson, F Zemrak, R Boubertakh, A A. Young, S Hudson, P Weale, S Garratt, R Collins, S Piechnik, and S Neubauer, "Uk biobank's cardiovascular magnetic resonance protocol," *Journal of Cardiovascular Magnetic Resonance*, 2016.

[10] I. Guyon and A. Elisseeff, "An introduction to variable and feature selection," *J. Mach. Learn. Res.*, vol. 3.